\documentclass{article}
\title{Statistics of the single mode light \\in the $\chi^{(3)}$
transparent medium}
\author{V.N.Gorbachev\footnote{email: vn@VG3025.spb.edu} 
and A.I.Zhiliba\footnote{email: Anatoly.Zhiliba@tversu.ru}
}

\begin{document}
\maketitle

{\small
\begin{center}
Department of Physics, University of Sankt-Petersburg,\\
Department of Physics, Tver State University*,\\
170006,Tver, Private Bag 0609$^\dag$, Russia
\end{center}
}
\medskip

\begin{abstract}
The quantum statistics of the light in the transparent medium with
cubic nonlinearity is considered. Two types of transparent media
are treated,namely, the cold transparent medium with a ground working
level and the inversion-free medium with the lasing levels of the same
population. The spectra of light fluctuation are found on the basis
of both Scully-Lamb and Haken theories. The conditions for the use of
effective Hamiltonian are determined.
Basing on the exact solution of the Fokker-Plank equation for the
Glauber-Sudarshan P-function the inversion-free medium with cubic
nonlinearity is shown to be the source of spontaneous radiation with
non-Gaussian statistics.
\end{abstract}

\section*{Inroduction}

Four-wave-mixing processes (FWM) are of particular interest for the study
of nonclassical states of the light as predicted by Yuen and
Shapiro \cite{H.P.Yuen1} and observed by Slusher et al. \cite{Slusher2}.
The quantum theory of FWM is based on two general approaches.
The first is the Effective Hamiltonian Approach (EHA) when a nonlinear medium
is described by a c-number coupling constant in an effective Hamiltonian of
interaction. The starting point of the second approach is the master equation
for the field obtained by adiabatic elimination of the atomic variables.
EHA has been used to consider the time evolution of the number of optical
systems \cite{J.Perina3}. Loudon et al have developed the quantum theory of
propagation of the light and examined squeezing produced by self-phase 
modulation in fiber and in the nonlinear Saqnac interferometer with the help 
of the effective Hamiltonian \cite{K.J.Blow4}. The effective Hamiltonians 
are usually introduced phenomenologically and occasionaly exact solution may 
be obtained. EHA seems to be attractive because it requires only 
susceptibilities to describe the nonlinear medium. For three-photon 
parametric phenomena the effective Hamiltonian can be obtained 
if the $\chi ^{(2)}$ medium is assumed to be transparent as it has been shown 
in \cite{Gorbachev6}. As to the case of $\chi ^{(3)}$
the above mentioned conditions \cite{Gorbachev6} for the use of effective 
Hamiltonian construction would not be expected due to the existence of real 
transitions.It is well known that the fluctuation-dissipation theorem (FDTh) 
does not exist in the case of four-index susceptibilities of a $\chi ^{(3)}$ 
medium \cite{Stratanovich5}. 

The second approach has been developed for two-level models: the Scully-Lamb
model (SLM) for the excited-state atom \cite{Sargent7} and the Haken model 
(HM) for the ground-lower level atom \cite{Bonifacio8},\cite{Reid9}.
Walls and Reid have shown that statistics of the light of nondegenerate FWM
under linear theory over two weak modes is the same in both HM and SLM
\cite{Reid9}. For the single-mode FWM the master equations of HM and SLM are 
markedly different and the two models are in agreement only for the
particular case of the linear theory \cite{Gorbachev6}.

The difference between HM and SLM arises not only from the type of the
two-level atom. Formalism of the Haken theory is based on the multi-atom
problem \cite{Haken10}. In contrast the starting points of the Scully-Lamb 
theory is the single-atom problem \cite{Scully11}. In this view these theories are
sometimes classified as the multi-atom and single-atom approaches \cite{Lugiato12}.
The problem is to find the most suitable approach \cite{Lugiato12}, \cite{Schack13}.
We believe that the HM and SLM describe an optical system in a quite 
different manner. It is an optical cavity with atoms inside for the HM while 
in the case of the SLM the atomic beam is injected into a cavity like in a 
maser or micromaser. It is clear that statistics of the light produced by the 
two systems may be substantially different. Here we use terms HM and LSM to 
take into account both the type of the approach (multi- or single-atom) and 
the type of the two-level atom. In the present paper we are aimed to apply 
our previous calculation within the framework of the Lax-Louisel model (LLM) 
to consider whether the effective Hamiltonians may be obtained for the $\chi^{(3)}$ 
medium and whether the statistics of the light is different for HM and LSM.

We discuss in this paper an optical scheme including the cavity with
nonlinear medium inside and the classical injected signal. The $\chi^{(3)}$
medium is described by the Lax-Louisell model which may be directly reduced
to the HM and SLM. Two kinds of transparent medium are considered. In
Section 1 spectra of fluctuation of the output light is calculated for the
cold-transparent medium for which only the lower working level is populated
and the frequency of the light is far from the atomic resonance. In Section
2 we consider the inversion-free medium when the population of both working
levels are equal. The semiclassical Maxwell theory predicts no evolution of
the light in the inversion-free medium. From this point of view the medium
is transparent. Nevertheless it is a source of the light analogous with that
of spontaneous radiation. With the use of the exact solution of the Fokker-
Planck equation for the Glauber-Sudarshan P-function the statistics of the
light is found to be non-Gaussian.

\section{ The cold transparent medium}

\subsection{Initial equations}

Let us consider a system consisting of $N$ the similar three-level fixed
atoms shown in Fig.1a. It is the Lax-Louisell model when the incoherent
pumping provides the populations $f^s_1$ and $f^s_2$ of working levels 1 
and 2, respectively, under conditions of the field absence. The medium are
placed into the high $Q$ cavity (see Fig.1b) at $\omega$ frequency, which can
differ from the injected field frequency $\omega_L$. The injected signal is
considered to be monochromatic coherent wave.
\begin{center}
\hspace {-1 cm} \unitlength=1cm
\begin{picture}(7,6)
\put(1,0){\line(1,0){5.5}}
\put(0.7,1.5){\line(1,0){3.7}}
\put(0.7,4.5){\line(1,0){5.5}}
\put(3,3.5){\line(1,0){1}}

\put(2.5,1.5){\vector(0,-1){1.5}}
\put(3,0){\vector(0,1){1.5}}
\put(5.2,4.5){\vector(0,-1){4.5}}
\put(5.8,0){\vector(0,1){4.5}}
\put(1.2,4.5){\vector(0,-1){3}}
\put(1.7,1.5){\vector(0,1){3}}
\put(3.5,1.5){\vector(0,1){2}}
\put(3.5,3.5){\vector(0,1){1}}
\put(3.5,4.5){\vector(0,-1){1}}

\put(2,0.7){$\gamma_1$}
\put(3.2,0.7){$\Lambda_1$}
\put(4.8,2.7){$\gamma_2$}
\put(5.9,2.7){$\Lambda_2$}
\put(3.7,2.7){$\omega_2$}
\put(0.7,3){$\gamma_{\downarrow}$}
\put(1.8,3){$\gamma_{\uparrow}$}
\put(3.7,4.05){$\Delta$}
\put(0.3,1.4){1}
\put(0.3,4.3){2}
\end{picture}
\\Fig. 1a. \\

\unitlength = 1cm
\begin{picture}(11,5)
\put(2.6,0.7){\line(1,0){2.5}}
\put(0.5, 3){\line(1,0){2.5}}
\put(4.6, 3){\line(1,0){2.2}}
\put(4.6, 3){\vector(1,0){0.5}}
\put(6, 3){\vector(1,0){0.4}}

\put(3,2.5){\framebox(1.55,1.2){$\chi^{(3)}$}}
\put(6.8, 2){\framebox(2.7,2){Homodyne Det}}

\put(0.4,3.1){$\omega_L$}
\put(6.2,3.1){$\omega$}
\put(1.3,3.7){$T^{in}$}
\put(5.7,3.7){$T^{out}$}
\put(3.5,1.7){$\omega$}

\put(3.5, 0.7){\vector(-1,1){2.25}}
\put(6,3){\vector(-1,-1){2.27}}

\put(1, 2.7){\line(1,1){0.55}}
\put(6.2, 2.7){\line(-1,1){0.55}}
\end{picture}
\\Fig.1b. \\
\end{center}

On the base of this model the master equation was derived by the adiabatic
elimination of atomic variables in \cite{Gorbachev6} where the field interaction 
with working transition was described by the Hamiltonian written in the dipole
and rotating-field approximation. The resulted equation was the
Fokker-Planck one for the Glauber-Sudarshan quasi-probability.
\begin{eqnarray}
\partial_t P=(\partial_\alpha\mu_1+\partial^2_{\alpha\alpha}\mu_2+
\partial^2_{\alpha\alpha^*}\mu_3+c.c.)P+R^{\prime}_FP,
\end{eqnarray}
where the coefficients $\mu_i$ (correct to $g^4$) for the medium with cubic
nonlinearity are:
\begin{eqnarray}
\mu_1=N\frac{g^2\delta^*}{|\delta|^2}\alpha(1-\beta|\alpha|^2)(f^s_1-f^s_2)
\nonumber \\
\mu_2=-N\alpha^2\frac{g^2\delta^*}{|\delta|^2} \lbrace\frac{g^2\delta^*}{%
|\delta|^2}(f^s_2-f^s_1) (\frac{1-f^s_1}{\Gamma_1}+\frac{f^s_2}{\Gamma_2}
\nonumber \\
+\frac{\delta^*}{|\delta|^2}(f^s_2-f^s_1) )+\frac{\beta\delta}{2\gamma}f^s_2]
\nonumber \\
\mu_3=-N\frac{g^2\delta^*}{|\delta|^2} \{\frac{g^2|\alpha|^2}{|\delta|^2}%
(f^s_2-f^s_1) \lbrace\frac{\delta}{\Gamma_2}f^s_2+\frac{\delta}{\Gamma_1}%
(1-f^s_1) \\
+f^s_2-f^s_1] +\beta|\alpha|^2\frac{\Gamma_1}{\Gamma_1+\Gamma_2}%
(f^s_2-f^s_1)+ \beta|\alpha|^2\frac{\delta^*}{2\gamma}f^s_2-f^s_2\}
\nonumber \\
\delta=\gamma+i\Delta  \nonumber \\
\Delta=\omega_{21}-\omega  \nonumber
\end{eqnarray}
$\gamma$ and $\omega_{21}$ are the constant of the transverse relaxation and
the frequency of the atomic transition respectively, $g=d\hbar^{-1}(\hbar%
\omega/2V\varepsilon _0)^{1/2}$ is the coupling constant, $d$ is the dipole
moment of the working transition, and $\beta=2g^2\gamma(\Gamma_1+\Gamma_2)(%
\Gamma_1\Gamma_2|\delta|^2)^{-1}$ is the saturation parameter. The relaxation
constants $\Gamma_1$and $\Gamma_2$ are formed from the set of initial
constants:
\begin{eqnarray}
\Gamma_2=(\gamma_\downarrow+\gamma^{\prime}_2)+(\gamma_\uparrow+\gamma^{%
\prime}_1) (\Lambda_1+\Lambda_2+\gamma_2)(\Lambda_1+\Lambda_2+\gamma_1)^{-1}
\nonumber \\
\gamma^{\prime}_k=\gamma_k(1-\Lambda_k(\Lambda_1+\Lambda_2)^{-1})  \nonumber
\\
k=1,2.  \nonumber
\end{eqnarray}
The constant $\Gamma_1$ can be obtained from $\Gamma_2$ by replacing . $%
\gamma_\uparrow \rightarrow \gamma_\downarrow $, $\gamma_\downarrow
\rightarrow \gamma_\uparrow $, $2 \rightarrow 1$, $1\rightarrow 2$. The term
\[
R^{^{\prime}}_F=\partial_\alpha(\frac12 C\alpha-a_0
\exp(i(\omega-\omega_L)t))+c.c.
\]
simulates the input and output of radiation from the cavity \cite{Scully11},\cite{Schack13}. 
Here $C=C_{in}+C_{out}$, $C_{in(out)}=c\,T^{in(out)}L^{-1}$, $c$ is the light
velocity, $T^{in(out)}$ is the transmission coefficient of the input and
output mirrors, $L$ is the cavity perimeter, and $a_0$ is the amplitude of the
injected field in inside the cavity. This amplitude is considered to be real.

Equation (1) describes the linear and nonlinear processes of interaction.
One can introduce the macroscopic susceptibilities for these processes.
\begin{eqnarray}
\chi ^{(1)}(\omega ) &=&Ng^2(\omega _{21}-\omega -i\gamma )^{-1}  \nonumber
\\
\chi ^{(3)}(\omega ) &=&\beta \chi ^{(1)}(\omega )
\end{eqnarray}
The Maxwell equation for the complex amplitude of the field inside the cavity $%
z=\int d^2\alpha \alpha P(\alpha ,t)$, follows from Eq.(1) Accounting for
Eq.(3) this equation can be written as
\begin{equation}
\partial _tz=((i\kappa _1-\kappa _2)z+(-i\chi _1+\chi _2)z|z|^2)(f_1^s-f_2^s)
\nonumber \\
-\frac C2z+a_0\exp (i(\omega -\omega _L)t)
\end{equation}
where the real and imaginary parts of the susceptibilities are separated out
\begin{eqnarray}
\chi ^{(1)} &=&\kappa _1+i\kappa _2  \nonumber \\
\chi ^{(3)} &=&\chi _1+i\chi _2  \nonumber
\end{eqnarray}
Since Eq. (1) is written correct up to $g^4$, the dimensionless intensity of
the field cannot be large.
\begin{equation}
\beta |z|^2,\quad \beta \langle |\alpha |^2\rangle \,\ll 1
\end{equation}
We consider the cold transparent medium assuming detuning $x=\Delta /\gamma $%
to be large.
\begin{equation}
|x|\gg 1
\end{equation}
Under these conditions the real and imaginary parts of the susceptibilities
are of essentially different order.
\begin{eqnarray}
\kappa _1 &\sim &Ng^2/\Delta ,\quad \chi _1\sim Nd^4/\Delta ^3  \nonumber \\
\kappa _2 &=&\kappa _1/x,\quad \chi _1=\chi _1/x  \nonumber
\end{eqnarray}
Therefore the linear and nonlinear dispersion described by the
susceptibilities $\kappa _1$and $\chi _1$ have a dominant role in a transparent
medium where $f_2^s=0$. The processes of absorption can be neglected setting
$\kappa _1,\chi _2=0$. However the last-mentioned fact means that the main
losses of the field in the cavity are connected only to the radiation
output, i.e., $\kappa _2,\,\chi _2|z|^2\ll C/2$.

As a result for the cold ($f_2^s=0$) transparent medium we have:
\begin{eqnarray}
\mu _1 &=&[-i\kappa _1\alpha +i\chi _1\alpha |\alpha |^2]f_1^s  \nonumber \\
\mu _2 &=&\alpha ^2\chi _1(1+\Gamma _1/\Gamma _2)^{-1}f_1^s(-\frac x2%
(1-f_1^s)+  \nonumber \\
&&+i(f_1^s-1-f_1^s\Gamma _1/2\gamma )) \\
\mu _3 &=&|\alpha |^2\chi _1(1+\Gamma _1/\Gamma _2)^{-1}f_1^s(1-f_1^s)\frac x%
2  \nonumber
\end{eqnarray}
The Lax-Louisell model allows a direct transition to HM and SLM. In \cite{Gorbachev6} 
the necessary conditions were formulated which can be reduced to the set
of inequalities between constants $\gamma _{\uparrow },\gamma _{\downarrow }$%
and $\gamma _1^{\prime },\gamma _2^{\prime }$. \newline
So, putting in Eq.(7)
\begin{eqnarray}
\Gamma _1 &=&\Gamma _2=\gamma _{\uparrow }+\gamma _{\downarrow }=\gamma
_{||},\quad f_1^s+f_2^s=1  \nonumber \\
\gamma &=&\gamma _{\perp }  \nonumber \\
\gamma _{\Vert }/2\gamma _{\perp } &=&f_c  \nonumber
\end{eqnarray}
we obtain the Fokker-Planck equations for HM.
\begin{equation}
\partial _tP=(\partial _\alpha (-i\kappa _1\alpha +i\chi _1\alpha |\alpha
|^2)  \nonumber \\
+\partial _{\alpha \alpha }^2(-\frac i2\chi _1\alpha
^2f_c)+c.c.)P+R_F^{\prime }P,
\end{equation}
In this equation the various types of relaxation are described $0<f_c\leq 1$,
the $f_c$ characterizes the difference between longitudinal and
transversal relaxation velocities. The case when $f_c=1$ describes the
radiation decay only. We note that Eq.(8) agrees with equation obtained in
\cite{Gardiner Handbook14}.\newline
Putting in Eq.(7)
\begin{eqnarray}
\Gamma _k &=&\gamma _k  \nonumber \\
(f_k^s)^2 &=&0,\quad f_k^s\ne 0,\quad k=1,2
\end{eqnarray}
we obtain the SLM equation.
\begin{eqnarray}
\partial _tP &=&f_1^s(\partial _\alpha (-i\kappa \alpha +i\chi \alpha
|\alpha |^2)  \nonumber \\
&&+\partial _{\alpha \alpha }^2(-\chi _1(1+\gamma _1/\gamma _2)^{-1}\alpha
^2(i-x/2)) \\
&&+\partial _{\alpha \alpha ^{*}}^2(\chi _1(1+\gamma _1/\gamma
_2)^{-1}|\alpha |^2x/2)+c.c.)P+R_F^{\prime }P  \nonumber
\end{eqnarray}
The conditions (9) include the requirement for weak pumping $(f_k^s)^2=0$,%
$f_k^s\ne 0$, when the terms proportional to $(f_k^s)^2$ are absent.
Formally, it is the consequence of the single-atom approach of the SLM.
Physically, SLM describes optical systems in which the atoms are
injected into the cavity being in a prepared state and rapidly escaped from
the interaction zone. The atoms output from the cavity is modeled by their
output from the working levels. On one hand, it requires an introduction of
two-level systems with excited levels, and, on the other hand, it allows the
use of the single-atom approach.

The Lax-Louisell model discussed here a priori describes the atom with two
excited working levels. As it follows from Eq.(7) at
\begin{eqnarray}
f^s_1=1,\quad \Gamma_1\Gamma_2((\Gamma_1+\Gamma_2)\gamma)^{-1}=f_c
\end{eqnarray}
the Fokker-Planck equations in the Haken and the Lax-Louisell models are the
same. It means that the statistics of the field formed by two-level
atoms with ground state and that of the field formed by atoms with two
excited levels is the same. Therefore under conditions (11) the differences
between LSM and LLM mainly depend upon the type of physical system rather
than the type of two-level atoms. We note that these differences disappear
when the linear over interaction theory is used. This theory is used as an
example in the case of the linear medium or in the case of multi-photon
interaction described by the effective Hamiltonian of the field in the
linear approximation \cite{Gardiner Handbook14}.

\subsection{The effective Hamiltonian}

The effective Hamiltonian introduced phenomenologically is often used to
describe the four-wave mixing in the transparent medium with  cubic
nonlinearity. The Hamiltonian is
\begin{eqnarray}
H_{eff}=\frac12 \hbar k {a^+}^2a^2,
\end{eqnarray}
where $k$ is the coupling constant. The exact equation for the
Glauber-Sudarshan quasi-probability written on the base of Eq.(12) has the
Fokker-Plank equation form
\begin{eqnarray}
\partial_tP=(\partial_\alpha(ik\alpha|\alpha|^2)+\partial_{\alpha \alpha}^2
(-\frac{i}{2}k\alpha^2)+c.c.)P.  \nonumber
\end{eqnarray}
For the cold transparent medium described by the Lax-Louisell model the
effective Hamiltonian appears solely in the case when
\begin{eqnarray}
f^s_1=1,  \nonumber \\
\Gamma_1\Gamma_2((\Gamma_1+\Gamma_2)\gamma)^{-1}=1.
\end{eqnarray}
In the HM it means that $f_c=1$, i.e., the decay is of the purely radiation
character. Under conditions of Eq.(13) the equation for the electromagnetic
field density matrix $\rho$ following from Eq.(7) or Eq.(8) takes the form
\begin{eqnarray}
\partial_t\rho=-i\hbar^{-1}[H_1+H_2,\rho]+R_F\rho  \nonumber \\
H_1=-\hbar\kappa_1 a^+a  \nonumber \\
H_2=\frac12 \hbar\chi_1 {a^+}^2a^2
\end{eqnarray}
where the operator $R_F\rho$ in $P$-representation corresponds to the $%
R^{\prime}_F\rho$ operator. In contrast to the effective Hamiltonian approach
(EHA) based on Eq.(12) the effect of the linear dispersion described by the
Hamiltonian $H_1$ is taken into account in Eq.14.

Physically speaking the linear dispersion tends to the change of mode
frequency $\omega\rightarrow\omega-\kappa_1$. It results the additional
detuning between the injected field frequency and the cavity mode. The role
of this factor was discussed in \cite{Reid9} and, as will be shown, it
substantially affects the fluctuation dynamics and the noises. If the
injected signal frequency is such that detuning is absent, i.e., $%
\omega_L-(\omega-\kappa_1)=0$, the linear dispersion is not important in the
transparent medium. In such a case the description by EHA with $%
k=Re\chi^{(3)}$ coincides with that by Eq.(14). The effective Hamiltonian
principally does not arise in the SLM due of the terms proportional to $x$.

\subsection{The Langevin equations}

The condition of the small fluctuations allowing the linearization of the
Fokker-Planck equation (FPE) is used as a basic approximation for the
calculation of observed values. Basing on the linearized equation  we write
the Langevin equations which are convenient to obtained the fluctuation
spectrum. We introduce polar coordinates
\[
\alpha \alpha^*=I,\quad \varphi=(1/2i)\ln(\alpha/ \alpha^*)
\]
The fluctuations of intensity and phase around its semi-classical values are
assumed to be small
\begin{eqnarray}
I=U+\varepsilon \qquad \varepsilon \ll U  \nonumber \\
\varphi=\varphi_0+\psi\qquad \psi\ll \varphi_0
\end{eqnarray}
where $U=|z|^2$ and $\varphi_0=\arg z$ are the steady-state solutions of the
semi-classical equations (4), which for transparent medium take the form  :
\begin{eqnarray}
\partial_tU&=-CU+2a_0\sqrt U \cos\varphi_0=0  \nonumber \\
\partial_t\varphi&=\kappa_1-\chi_1U+a_0\sin\varphi_0/\sqrt U=0
\end{eqnarray}
In Eqs.( 16 ) we assumed the phase of the injected field to be equal to zero
and $\omega_L=\omega$. As a result we obtain the Langevin equation
\begin{eqnarray}
\partial_t \left( \matrix{\varepsilon \cr \psi} \right) = \left( \matrix{
-A,&A_{12}\cr A_{21},&-A} \right) \left( \matrix{\varepsilon \cr \psi}
\right) + D^{1/2}\eta(t)
\end{eqnarray}
where $\eta(t)$ is the two-dimensional Winer process and the elements of the
drift $A$ and diffusion $D$ matrix for the three considered models are listed
in Table 1. Note that the linear dispersion affects only the fluctuations
dynamics, and in SLM the source of the phase noise appears.

\begin{center}

Table 1.
\begin{tabular}{|c|c|c|c|}
\hline
& EHA & HM & SLM \\
\hline
A & \multicolumn{3}{|c|}{C/2} \\ \hline
$A_{12}$ & $2kU^2$ & \multicolumn{2}{|c|}{$2\chi_1 U^2-2U \kappa_1$} \\
\hline
$A_{21}$ & $-\frac{3}{2}k$ & \multicolumn{2}{|c|}{$-\frac{3}{2}\chi_1 +\frac{%
1}{2}\kappa_1/U$} \\ \hline
$D_{\epsilon\phi}$ & $-kU$ & $-\chi_1Uf_c$ & $-\chi_1U$ \\ \hline
$D_{\psi\psi}$ & \multicolumn{2}{|c|}{0} & $\frac{1}{2}x\chi_1$ \\ \hline
\end{tabular}
\end{center}

The phase and the intensity fluctuation are coupled dynamically $%
A_{12},A_{21}\ne 0$ and statistically $D_{\varepsilon \psi }\ne 0$. This fact
leads to two circumstances. First, both the intensity and phase fluctuations
is produced by two sources $D_{\varepsilon \psi }$and $D_{\psi \psi }$.
Second, under steady-state the fluctuations being attenuated oscillate with
the frequency $\Omega $. To obtain $\Omega $ we write closed equations for
the averaged values. These equations follow from Eq.(17).
\begin{eqnarray}
(\partial _t^2+2A\partial _t+A^2+\Omega ^2)\sigma  &=&0  \nonumber \\
\sigma  &=&\langle \varepsilon \rangle ,\langle \psi \rangle  \\
\Omega ^2 &=&-A_{12}A_{21}  \nonumber
\end{eqnarray}
In HM and SLM
\[
\Omega ^2=\kappa ^2(1-\beta U)(1-3\beta U),
\]
where the dimensionless intensity $\beta U\ll 1$ since the initial equation
(1) is correct up to $g^4$. Let
\begin{eqnarray}
\beta U\ll 1/3
\end{eqnarray}
Then $\Omega ^2=\kappa ^2$and $\Omega $ takes a new meaning of the
damped harmonic oscillator frequency described by Eq.(18). As a result under
steady-state the fluctuations of phase and intensity damp with velocity $%
C/2$ and oscillate at the frequency $\Omega $.

In HM and SLM under conditions (19) the dynamics of the fluctuations are
defined solely by the linear dispersion.
\[
A_{12}=-2U\kappa _1,\qquad A_{21}=\frac 12\kappa _1/U
\]
where the expression for $U$ follows from Eqs.(16) under conditions (19).
\begin{eqnarray}
U=a_0^2(A^2+\kappa _1^2)^{-1}
\end{eqnarray}
The output power $P=\hbar \omega cUT^{out}L^{-1}$ will be connected
to the power of the injected light $P_0$ with the use of Eq.(20) as
\[
P=P_0\frac 1{1+t^2}\cdot \frac{4\epsilon }{(1+\epsilon )^2}
\]
where
\begin{eqnarray}
t=2\kappa _1/C,\qquad \epsilon =T^{in}/T^{out}
\end{eqnarray}

\subsection{The fluctuations spectra}

For the homodyne balanced detection, when the signal is mixed with the field
of local oscillator, the normal-ordered correlation functions of quadrature
operator
\begin{eqnarray}
X(t,\Theta )=a^{+}(t)\exp (i\Theta )+h.c.
\end{eqnarray}
are measured. The spectrum of the photocurrent $i^{(2)}(\omega )$ or the noise
power spectrum can be presented in the form:
\begin{eqnarray}
i^{(2)}(\omega ) &=&1+\eta g(\omega ,\Theta )  \nonumber \\
g(\omega ,\Theta ) &=&C^{out}\int_{-\infty }^{+\infty }d\tau e^{i\omega \tau
}\langle :X(t,\Theta )X(t+\tau ,\Theta ):\rangle
\end{eqnarray}
where $\eta $ is the quantum efficiency of the recording scheme, the points
denote the normal ordering, and scale is determined so that the unit
corresponds to the shot noise level. The value $g(\omega ,\Theta )$can be
calculated by introduction of the Langevin variable
\begin{eqnarray}
X(t,\Theta )=\alpha ^{*}(t)\exp (i\Theta )+c.c.
\end{eqnarray}
which is $c$-number analog of the operator (22). So, it is easy to show
that
\begin{eqnarray}
g(\omega ,\Theta )=C^{out}\int_{-\infty }^{+\infty }d\tau e^{i\omega \tau
}\langle X(t,\Theta )X(t+\tau ,\Theta )\rangle
\end{eqnarray}
The Eq. (25) is true when the Langevin equations are written from the FPE
for the Glauber P-function or, as an example, the generalized
P-representation. With the use of the small fluctuations (15)
\[
X(t,\Theta )=2\sqrt{U}\cos \varphi _0+2\sqrt{U}\sin (\Theta -\varphi _0)\psi
(t)\\+\frac 1{\sqrt{U}}\cos (\Theta -\varphi _0)\varepsilon (t)
\]
Therefore the case when the phase of the local oscillator field coincides
with that of the signal, i.e. $\Theta =\varphi _0$, corresponds to the
measurement of spectrum of the intensity fluctuations, while when $\Theta
-\varphi _0=\frac \pi 2$ the phase fluctuations are measured. Solving the
Langevin equation (17) we obtain
\begin{eqnarray}
g(\omega ,\Theta ) &=&2C^{out}[(\omega ^2+S_1^2)(\omega ^2+S_2^2)]^{-1}\cdot
\nonumber \\
&&\cdot (D_{\psi \psi }[(\mu A+\nu A_{12})^2+\mu ^2\omega ^2]  \nonumber \\
&&+2D_{\epsilon \psi }[(\mu A+\nu A_{12})(\mu A_{21}+\nu A)+\mu \nu \omega
^2]  \nonumber
\end{eqnarray}
where $\mu =\sqrt{2U}\sin {(\Theta -\varphi _0)}$, $\nu =\frac 1{\sqrt{2U}}%
\cos {(\Theta -\varphi _0)}$, and $S_{1,2}$ are the roots of the equation $%
(S+A)^2-A_{12}A_{21}=0$. Introducing the scaled frequency $\bar \omega
=\omega /A$, $t=\kappa _1/A$ we represent the obtained spectrum in the form
\[
g(\omega ,\Theta )=\frac 4{1+\epsilon }\frac{W(\Theta -\varphi _0)\cdot
t^2+V(\Theta -\varphi _0)\cdot {\bar \omega }^2}{({\bar \omega }%
^2+1-t^2)^2+4t^2}
\]
For the measurement of the intensity and the phase fluctuations the coefficients
are presented in Table 2. For SLM the case $\gamma _1=\gamma _2$, $f_1^s=1$%
is taken as before.

\begin{center}
Table 2.
\begin{tabular}{|c|c|c|c|}
\hline
& EHA & HM & LSM \\ \hline
$W(0)$ & -2/3 & $2\beta Uf_{c}$ & $2\beta U(1+\frac{xt}{2})$ \\ \hline
$V(0)$ & \multicolumn{3}{|c|}{$0$} \\ \hline
$W(\frac{\pi}{2})$ & $2$ & $-2\beta Uf_{c}$ & $-2\beta U(1-\frac{x}{2t})$ \\
\hline
$V(\frac{\pi}{2})$ & \multicolumn{2}{|c|}{$0$} & $\beta Uxt$ \\ \hline
$t$ & $\sqrt{3}kU/A$ & \multicolumn{2}{|c|}{$\kappa_1/A$} \\ \hline
\end{tabular}
\end{center}

When EHM and HM are compared one can see that the effect of dispersion is
diametrically opposite behavior both for the phase and amplitude fluctuations.
However on the basis of both approaches compaction arises
leading to a decrease in the shot noise by a factor of not more than 3.
Indeed the noise reduction may be obtained in nonzero frequency range.
This behavior is understood by suggestion that the
dispersion rotates the squeezing ellipse.

To determine the degree of squeezing for the optical scheme under
consideration we need to choose the optimal phase of the local oscillator $%
\Theta=\Theta_0$. To this end we represent $g(\omega,\Theta)$ in the form $%
g=A+B\sin2\Theta+C\cos2\Theta$ and determine the $\Theta_0$ value from the
condition $\tan 2\Theta_0=B/C$. As a result the fluctuation spectrum takes a
form\\
\begin{eqnarray}
g(\omega,\Theta_0)=\frac{2}{1+\epsilon}[({\bar\omega}%
^2+1-t^2)^2+4t^2]^{-1}G(\omega)  \nonumber
\end{eqnarray}
where\\
\begin{eqnarray}
G_{EHA}(\omega)=\frac{2t}{3}(2t\pm(4t^2+3(1-t^2+{\bar\omega}^2)^2)^{1/2})
\nonumber \\
G_{HM}(\omega)=\pm2f_c\beta u|t|\cdot M  \nonumber \\
G_{LSM}(\omega)=2\beta U x t [x(1+{\bar\omega}^2+t^2)\pm M]  \nonumber \\
M=[(1+t^2)^{-1}[(1+{\bar\omega}^2+t^2)^2+t^2(1+t^2-{\bar\omega}^2)^2]]^{1/2}
\nonumber
\end{eqnarray}
In the zero-frequency range
\begin{eqnarray}
g_{EHA}(0,\Theta)=\frac{4}{3}\frac{t}{(1+t^2)^2}(2t\pm\sqrt{4t^2+3(t^2-1)^2})
\nonumber \\
g_{HM}(0,\Theta)=\pm4 f_c\beta U\frac{|t|}{|t|^2+1}\qquad {squeezing} -2/3
\nonumber \\
g_{LSM}(0,\Theta)=\pm \frac{4}{x}\beta U\frac{t}{t^2+1}\qquad {weak squeezing%
}  \nonumber
\end{eqnarray}
For EHM the noise can be suppressed not only in the zero-frequency range.
Choosing the local oscillator field phase using the condition $\tan\Theta=(%
\sqrt 3 t)^{-1}$, we find\\
\begin{eqnarray}
g_{EHA}(\omega)=-\frac{8{\bar\omega}^2}{C[({\bar\omega}^2+1-t^2)^2+4t^2]}%
\cdot \frac{t^2}{1+3t^2}.  \nonumber
\end{eqnarray}

At the frequency ${\bar\omega_e}^2=1+t^2$ we find the minimum\\

\[
g_{min}({\bar\omega_e})=-\frac{2t^2}{C(1+3t^2)}\\
\]\\
The maximum drop is equal 2/3.

\section{ Inversion-free transparent media}

\subsection{ Initial equations}

Let the pumping produce equal populations of working levels
$f_1^s=f_2^s=f_0$. As it seen from FPE (1), in such a medium there is no
polarization proportional to inversion. In this sense the inversion-free medium
is transparent. Nevertheless the field can evaluate in such a media,especially
in nonlinear one. At the same time it is seen directly from Eq.(1) that the
contribution of second derivatives formed from the terms of $\mu _2$ and $\mu
_3$ remains depending solely on the population of upper level $f_2^s$ so
the field can nevertheless evolute in such mediua, especially
in nonlinear ones. Therefore the field in a inversion-free medium has
a spontaneous radiation nature.

However the Eq.(1) does not permit the correct consideration of the field in
a inversion-free medium because of the procedure of its derivation. The reason
is that the noncommutativity of differential operators is neglected when the
second derivatives are obtained. It has been shown in \cite{Gorbachev6} that the
noncommutativity contributes to the first derivatives in FPE that correspond
to the semiclassical Maxwell's  equations.  If $f_1^s\ne f_2^s$ this
contributions can be neglected. However, for inversion-free medium the first
derivatives are produced by the noncommutativity only that why it must be
taken into account. Using the formalism from \cite{Gorbachev6} it is possible to
obtain the master equation for the field with due regard for
noncommutativity of differential operators. The Glauber-Sudarshan
quasi-probability equation correct to $g^4$ for LLM has the form 
\begin{eqnarray}
\partial _tP &=&\frac 12f_0(\frac \partial {\partial \alpha }\alpha \chi
_2[3+x^2-2ix]-\frac{\partial ^2}{\partial \alpha ^2}\alpha ^2\chi _2^2(1+x^2)
\nonumber \\
&&+\frac{\partial ^2}{\partial \alpha \partial \alpha ^{*}}[2\kappa _2+\chi
_2((x^2-1)|\alpha |^2-1+\frac{2f_0(\Gamma _2-\Gamma _1)-\Gamma _2}{\Gamma
_1+\Gamma _2})] \\
&&+\frac{\partial ^3}{\partial \alpha ^{*}\partial \alpha ^2}\alpha \cdot
2\chi _2[1-f_0\frac{\Gamma _2-\Gamma _1}{\Gamma _1+\Gamma _2}+ix\frac{%
(f_0-1)(\Gamma _2-\Gamma _1)}{\Gamma _1+\Gamma _2}]  \nonumber \\
&&+\frac{\partial ^4}{\partial \alpha ^2\partial {\alpha ^{*}}^2}2\chi _2f_0%
\frac{\Gamma _1}{\Gamma _1+\Gamma _2}+c.c)P  \nonumber
\end{eqnarray}
The polarization proportional to cubic susceptibility $\chi _2$ appearing in
inversion-free medium leads to evolution of the complex field amplitude $%
\langle \alpha \rangle =\int d^2\alpha \alpha P(\alpha ,t)$\newline
So, according to Eq.(26)
\[
\partial _t\langle \alpha \rangle =-\frac 12f_0[\chi _2(3+x^2)-2i\chi
_1]\langle \alpha \rangle
\]
The field amplitude damping and its steady-state value 
equals zero because $\chi _2=0$. From the physicall point of view this result 
is related to the diffusion of the spontaneous radiation phase. Equation (26) 
can be approximated by the following FPE 
\begin{eqnarray}
\partial _tP &=&\frac 12f_0(\frac \partial {\partial \alpha }\alpha \chi
_2[3+x^2-2ix]-\frac{\partial ^2}{\partial \alpha ^2}\alpha ^2\kappa _2(1+x^2)
\\
&&+\frac{\partial ^2}{\partial \alpha \partial \alpha ^{*}}[2\kappa +\chi
_2((x^2-1)|\alpha |^2-\frac 32)]+c.c)P+R_F^{\prime }P  \nonumber
\end{eqnarray}
where we set $\Gamma _1=\Gamma _2$ and add the term $R_F^{\prime }P$ assuming
that the medium is placed in the cavity with the injected signal on. The FPE
obtained can be solved exactly.

\subsection{The exact solutions for free-inversion transparent medium}

Let us represent the FPE (27) in the form
\begin{eqnarray}
\partial_t P=-\partial_\alpha(k\alpha+a_0)+{\partial_{\alpha\alpha}}%
^2\alpha^2 \Lambda_{\alpha\alpha} \\
+\partial^2_{\alpha\alpha^*}(b+|\alpha|^2\Lambda_{\alpha\alpha^*})+c.c)P
\nonumber
\end{eqnarray}
where $a_0$ and $b$ are the real constants. To obtain the solutions the
corresponding Langevin equations are used
\begin{eqnarray}
\partial_t\alpha=k\alpha+a_0+f_\alpha  \nonumber \\
\partial_t\alpha^*=(k\alpha+a_0)^*+f_{\alpha^*}
\end{eqnarray}
where nonzero correlation functions for random forces have the form
\begin{eqnarray}
\langle f_\alpha(t)f_\alpha(t+\tau)\rangle =2\langle \alpha^2\rangle
\Lambda_{\alpha\alpha}\delta(\tau)  \nonumber \\
\langle f_{\alpha^*}(t)f_{\alpha^*}(t+\tau)\rangle =2\langle {\alpha^*}%
^2\rangle \Lambda^*_{\alpha\alpha}\delta(\tau)  \nonumber \\
\langle f_{\alpha}(t)f_{\alpha^*}(t+\tau)\rangle =2(b+\langle
|\alpha|^2\rangle \Lambda_{\alpha\alpha^*}\delta(\tau)  \nonumber
\end{eqnarray}
The formulated Langevin approach  based on the FPE for the
Glauber-Sudarshan quasi-probability was clasified in \cite{Gardiner Handbook14}
as a "naive" one, because, for instance, the variables $\alpha$and $\alpha^*$
cannot be interpreted as independent ones. Here we do not discuss neither the
interpretation problems nor the mathematical aspects of stochastic differential
equation theory, but only notice the following. First, the Langevin approach
is in full accordance with the FPE (27) as far as concerned the
calculation of observed values and based on Itto interpretation of
stochastic integral. Second, all the results obtained below can be readily
reproduced using, for example, the generalized quasi-probability $%
P(\alpha,\alpha^+)$, where $\alpha$and $\alpha^+$ are independent variables
owing to the simple correspondence
\[
P(\alpha,\alpha^*)\longleftrightarrow P(\alpha,\alpha^+)\qquad
\alpha,\alpha^*\longleftrightarrow \alpha,\alpha^+
\]
Solving Eq.(29) we find
\begin{eqnarray}
\alpha(t)=z(t)+\int_0^t ds e^{k(t-s)}f_\alpha(s)  \nonumber \\
z(t)=\alpha(0)e^{kt}+\frac{a_0}{k}(e^{kt}-1)
\end{eqnarray}
The expression for steady-state of complex amplitude of the field in the
cavity can be consequently obtained.
\begin{eqnarray}
\langle \alpha\rangle =-a_0/k  \nonumber \\
Re k<0  \nonumber
\end{eqnarray}
Using Eq.(30) we find the correlation function
\begin{eqnarray}
Y(\tau,\Theta)=\langle X(t,\Theta)X(t+\tau,\Theta)\rangle ,\quad \tau>0
\nonumber \\
X(t,\Theta)=\alpha^*\exp(i\Theta)+c.c.  \nonumber
\end{eqnarray}
To do this we need in average values of the type $\langle
\alpha(t)\alpha(t+\tau)\rangle $, $\langle \alpha^*(t)\alpha(t+\tau)\rangle $%
and so on. For example, using Eq.(30) we obtain:
\begin{eqnarray}
\langle \alpha(t)\alpha(t+\tau)\rangle = z(t)z(t+\tau)  \nonumber \\
+e^{kt}[\langle \alpha^2(t)\rangle -z^2(t)+e^{2kt}\int_0^t ds_1\int_0^t ds_2
\nonumber \\
e^{-k(s_1+s_2)}\langle f_\alpha(s_1)f_\alpha(s_2)\rangle ]  \nonumber
\end{eqnarray}
Due to the delta-function properties this integral is equal to zero,
consequently, at $t\rightarrow \infty$
\[
\langle \alpha(t)\alpha(t+\tau)\rangle =\langle \alpha\rangle
^2+e^{k\tau}(\langle \alpha^2\rangle -\langle \alpha\rangle ^2)
\]
As a result at $t\rightarrow \infty$ the expression for $Y$ takes the form
\begin{eqnarray}
Y(t,\Theta)=\langle \alpha\rangle \langle \alpha^*\rangle
+e^{-2i\Theta}\langle \alpha\rangle ^2  \nonumber \\
+e^{kt}(\langle |\alpha|^2\rangle -\langle \alpha\rangle \langle
\alpha^*\rangle  \nonumber \\
+e^{2i\Theta}(\langle \alpha^2\rangle -\langle \alpha\rangle ^2))+c.c.
\end{eqnarray}
The expression for the Fourier transformation is written as
\begin{eqnarray}
Y(\omega,\Theta)=2{Re}\int_0^\infty d\tau e^{i\omega\tau}Y(\tau,\Theta)
\nonumber \\
Y(\omega,\Theta)=\frac{4}{A}[({\bar\omega}^2+1-T^2)^2+4T^2]^{-1}\cdot
\nonumber \\
\cdot((S+{Re}\mu)(1+{\bar\omega}^2+T^2)+T Im\mu({\bar\omega}^2-1-T^2))
\end{eqnarray}
where
\begin{eqnarray}
T={Im}\,k/A  \nonumber \\
\bar\omega=\omega/A  \nonumber \\
A=-{Re}\,k  \nonumber \\
S=\langle |\alpha|^2\rangle -\langle \alpha\rangle \langle \alpha^*\rangle
\nonumber \\
\mu=\exp(-2i\Theta)(\langle \alpha^2\rangle -\langle \alpha\rangle ^2)
\nonumber
\end{eqnarray}
It should be noted that Eq.(31) is valid for any diffusion coefficient of
FPE (28). Therefore the main problem reduces to the determination of the
steady-state variance $S$ and $\mu$.

Using Eq.(28) or the Langevin equations (29) the closed equations for $S$ and
$\mu $ can be obtained, which can be solved exactly. As a result, at
\begin{eqnarray}
{Re}\,k &<&0  \nonumber \\
{Re}\,(k+\Lambda _{\alpha \alpha ^{*}}) &<&0  \nonumber \\
{Re}\,(k+\Lambda _{\alpha \alpha }) &<&0  \nonumber
\end{eqnarray}
we obtain the necessary steady-state solutions:
\begin{eqnarray}
\langle |\alpha |^2\rangle  &=&(\frac{a_0^2}{|k|^2}(k+k^{*})-2b)(k+k^{*}+2%
\Lambda _{\alpha \alpha ^{*}})^{-1}  \nonumber \\
S &=&-2(\frac{a_0^2}{|k|^2}\Lambda _{\alpha \alpha ^{*}}+b)(k+k^{*}+2\Lambda
_{\alpha \alpha ^{*}})^{-1}  \nonumber \\
\mu  &=&-e^{2i\Theta }\frac{a_0^2}{|k|^2}\cdot \frac{\Lambda _{\alpha \alpha
}}{k+\Lambda _{\alpha \alpha }}
\end{eqnarray}
It is interesting to note the following fact. The Fokker-Planck equations
(8) and (9) for the SLM and the HM models reduce to Eq.(28) if the nonlinear
dispersion is neglected. For example, assuming that $\kappa \gg \chi
_1|\alpha |^2$, i.e., $\beta |\alpha |^2\ll 1$ we obtain in Eq.(28)
\begin{eqnarray}
k &=&-C/2+i\kappa _1  \nonumber \\
\Lambda _{\alpha \alpha } &=&(-\frac i2\chi _1f_c+\chi _1x/4)f_1^s  \nonumber
\\
\Lambda _{\alpha \alpha ^{*}} &=&f_1^sx\chi _1/4  \nonumber \\
b &=&0  \nonumber
\end{eqnarray}
where  $f_1^s=1$, $x=0$ should be set for the Haken model and $x\ne 0$, $%
f_c=1$ for the Scully-Lamb one. The saturation parameter $\beta $ can be
considered in transparent medium as a small one (in the sense of the
conditions (6)). This permits the neglect of $\Lambda _{\alpha \alpha }$and $%
\Lambda _{\alpha \alpha ^{*}}$ values in denominators of Eq.(33) when
calculating $S$ and $\mu $. This approximation properly implies a
linearization of the FPE near the steady-state $\langle \alpha \rangle $%
obtained according to Eq.(30). As a result the solutions obtained in Section
1 both for the Haken and the Scully-Lamb models can be found using the
presented formulas.

\subsection{Fluctuation spectra}

The term $\frac 12f_0\frac \partial {\partial \alpha }\alpha \chi _2(-2ix)$%
in FPE (27) corresponds to the dispersion leading to the change of cavity
mode frequency if $\omega \rightarrow \omega ^{\prime }-f_0\chi _2\cdot x$.
However, if the injected signal frequency is assumed to be $\omega _L=\omega
^{\prime }$ the dispersion plays no role. For this case the Eqs.(33) take the
form
\begin{eqnarray}
\langle |\alpha |^2\rangle  &=&(q_0+\frac{a_0^2}{|k|^2}(1+\frac 12\beta
q_0(3+x^2))R  \nonumber \\
S &=&(q_0-\frac{\beta a_0^2}{2|k|^2}q_0(1-x^2))R \\
\mu  &=&-\frac{\beta a_0^2}{2|k|^2}q_0(1+x^2)R\exp (-2i\Theta )  \nonumber \\
R &=&(1+2\beta q_0)^{-1}  \nonumber
\end{eqnarray}
where the assumption $q_0=2f_0\kappa _2/C$ is used and $1-3\beta /4\approx 1$
Because the value $k=-\frac C2[1+\frac 12\beta q_0(3+x^2)]$ is real the
expression for spectrum $Y(\omega ,\Theta )$ obtained according to Eq.(32)
takes the simple form
\begin{eqnarray}
Y(\omega ,\Theta )=\frac{8(S+{Re}\mu )}{C[1+\frac 12\beta q_0(3+x^2)](1+{%
\bar \omega }^2)}  \nonumber
\end{eqnarray}
Both the maximum and minimum values of $Y$ is obtained at $\Theta =\frac \pi 2%
$(max)

\begin{eqnarray}
S+{Re}\mu=q_0(1+x^2\beta a_0^2/k^2)(1+2\beta q_0)^{-1}  \nonumber
\end{eqnarray}
and $\Theta=0$ (min)

\begin{eqnarray}
S+{Re}\mu=q_0(1-\beta a_0^2/k^2)(1+2\beta q_0)^{-1}  \nonumber
\end{eqnarray}.
For the case $\Theta=0$, which corresponds to the measurement of
amplitude fluctuations, the value $S+{Re}\mu>0$. It follows
from the fact that initial equation is correct to the order of $g^4$
i.e., only the cubic nonlinearity is taken into account. Therefore it is
necessary to consider the field as not very strong one, i.e.,
\[
\beta a_0^2/k^2=\beta\langle \alpha\rangle \langle \alpha^*\rangle <1,\quad
\beta<|\alpha|^2\rangle <1
\]
As a result the noise is found to be greater that the shot one. Physically
it is clear because the inversion-free medium is the source of spontaneous
radiation, the statistics of which, however, can be non-Gaussian.

\subsection{ Generation of light by the inversion-free medium}

Let the injected signal and detuning be equal to zero. In such a
medium the field with average number of photons
\begin{eqnarray}
\langle n\rangle  &=&q_0(1+2\beta q_0)^{-1}  \nonumber \\
q_0 &=&2f_0\kappa _2/C
\end{eqnarray}
and the Lorentzian profile of spectral line with halfwidth
\[
\Delta \nu =\frac C2(1+\frac 32\beta q_0)
\]
is generated. Notice that the threshold of the generation is absent, and
dimensionless intensity is limited:
\[
\beta \langle n\rangle \leq 1/2
\]
Consider the statistics of the generated light. Using Eq.(27) the Mandel
parameter can be obtained:
\[
\xi =\langle n\rangle (1-\beta \langle n\rangle )\frac{1+2\beta q_0}{1+\frac
52\beta q_0}
\]
For the linear inversion-free medium $\beta =0$, $\xi =\langle n\rangle $%
i.e., the field statistics is of the Gaussian type. By contrast in the
case under consideration $\frac 25\langle n\rangle \leq \xi <\langle
n\rangle $. It means that the function of the photons number distribution is
narrower than the Gaussian one. Therefore the spontaneous radiation statistics
of the non-linear inversion-free medium is of essentially non-Gaussian nature.

\section{Conclusion}

The presented study of quantum statistics of single mode light in transparent $%
\chi {(3)}$ medium allows us to  compare the intensity and phase
fluctuation spectra obtained on the basis of both Haken and Scully-Lamb models.
The performed calculations clearly show  how the light
statistics in transparent $\chi {(3)}$ medium depends on the model
medium excitation. It is seen that the light in Scully-Lamb model is more noisy
than in the Haken one.
Physically, it is rather evidently since in the SLM the medium subsystem is open.
The atoms are injected into the cavity and leave the interaction zone rapidly.
As a result, the SLM the medium excitation has additional source
of noise whith relation to the Haken model.It is shown in this paper that
there are no any cases when EHA is valid in framework of Scully-Lamb model.
But for the cold transparent medium Haken model with the lower working level
to be ground state the traditional effective Hamiltonian of interaction was derived
in particular case of absence of the collision-induced  phase decay of atoms.

In the interesting case of the inversion-free $\chi {(3)}$ medium presented
study leads to description of quantum evolution of light based on reduction
of the Fokker-Plank equation from \cite{Gorbachev6} to equation (20) presented paper.
The obtained FPA has an exact solution. The statistics of generated light in
this case is nongaussian that is photon number distribution is narrower than
in the Gaussian statistics. At the same time there is no squeezing. Is
this state of generated light quantum because this state and its evolution
appear only in the framework of quantum description? It necessary that futher
study of the problem should be performed.\newpage


\end{document}